\documentclass[pra,aps,twocolumn,groupedaddress,superscriptaddress,floatfix]{revtex4}
\usepackage{epsfig}
\usepackage{amsmath}
\newcommand{\be}{\begin{equation}}
\newcommand{\ee}{\end{equation}}
\newcommand{\bea}{\begin{eqnarray}}
\newcommand{\eea}{\end{eqnarray}}

\newcommand{\bc}{\begin{center}}
\newcommand{\ec}{\end{center}}

\renewcommand{\(}{\left(}
\renewcommand{\)}{\right)}

\newcommand{\forget}[1]{}
\newcommand{\half}{\frac{1}{2}}

\begin{document}

\preprint{}
\title{Quantum key 
distribution without 
reference frame alignment: Exploiting photon orbital angular momentum }
\author{Federico M. Spedalieri}
\email{Federico.Spedalieri@jpl.nasa.gov}
\affiliation{Quantum Computing Technologies Group, Jet Propulsion Laboratory,
California Institute of Technology, Mail Stop 126-347, 4800 Oak Grove Drive, 
Pasadena, California 91109-8099}

\date{\today}

\begin{abstract}
We present a new implementation of the BB84 quantum key distribution protocol
that employs a $d$-dimensional Hilbert space spanned by spatial modes of
the propagating beam that have a definite value of orbital angular 
momentum. Each photon carries $\log d$ bits of information,
increasing the key generation rate of the protocol. The states used
in the transmission part of the protocol are invariant 
under rotations about the propagation direction, making this implementation 
independent of
the alignment between the reference frames of the sender and receiver.
The protocol still works when these reference frames rotate with 
respect to each other.

\end{abstract}
\maketitle

Quantum key distribution (QKD) is one of the most developed applications
of quantum information theory (QIT)~\cite{nielsen2000}. It is based 
on the properties 
of quantum states and allows two spatially separated parties to generate
a shared secret key. The key generated is secure in the sense
that an eavesdropper cannot obtain more than an exponentially small amount
of information about the key without being detected. 

Photons have been the information carriers of choice for quantum
key distribution. Photons can be sent through optical fibers for
relatively long distances (the main obstacle being photon absorption),
and they can also be sent through open air. A qubit (or bit) of information
can be encoded in the polarization degrees of freedom, and the linear
superpositions required by QKD can be easily produced with polarization
rotators. 

However, QKD protocols employing photon polarization 
have two main restrictions: they only allow transmission of 
one key bit per photon, and they require the reference frames
of the sender and receiver (usually known as Alice and Bob)
to be aligned with each other. The latter should be considered an
extra resource required by the protocol. It may not seem
too strong of a restriction for ground-based stations, but 
it is important if either Alice or Bob (or both) 
are based on a moving station such as a satellite. In this case
they must continually monitor and control this alignment.

A possible approach to achieve transmission of more than one
bit of information per photon is to employ orbital angular 
momentum (OAM) states of photons, since the Hilbert space spanned by these
states is in principle infinite. A lot of attention has been devoted
recently to the study of properties of OAM states and to
their generation and manipulation~\cite{molina2002a,
mair2001a,langford2004a}. On the other hand, 
it was shown in~\cite{bartlett2003a}
that classical and quantum information can be transmitted 
without a shared reference frame, but the implementation suggested
requires entangled states.  

In this letter, we present an implementation of the well-known 
BB84 protocol~\cite{bennett1984a} for QKD 
that goes beyond these two restrictions.
The protocol encodes the information 
in different spatial modes of the propagating photon that have
a definite value of OAM. By choosing
a subset of these modes we can effectively increase the amount of 
information encoded in each
photon. Furthermore, since these states are eigenstates of orbital
angular momentum, they are invariant under rotations
about the propagation direction of the beam. The QKD protocol
can be implemented without alignment of reference frames, and
without requiring entangled states of photons. 

We can write the electromagnetic vector potential for a linearly
polarized laser in the Lorentz gauge propagating in the 
$\hat{\mathbf{z}}$ direction as $\vec{A} = \hat{\mathbf{x}}\,
u(x,y,z)\, e^{-i k z }$. The spatial modes $u(x,y,z)$ can be obtained 
by solving
the wave equation for this particular ansatz. In the paraxial approximation 
there
are two important families of solutions, known as the Hermite-Gauss (HG)
modes, and the Laguerre-Gauss (LG) modes~\cite{beijersbergen1993a,
wunsche2004a,abramochkin2004a}. The respective spatial mode 
functions are given by
\bea
\label{modes}
u_{n m}^{HG} (x,y,z) & = & C_{n m}^{HG} \(\frac{1}{w}\) e^{-i [k \frac{
(x^2 + y^2)}{2 R} +(n+m+1) \psi]} \times \nonumber \\
& & \times e^{ - \frac{(x^2 + y^2)}{w^2}}
H_n\(\frac{x\sqrt{2}}{w}\) H_m\(\frac{y\sqrt{2}}{w}\), \\
\label{modesLG}
u_{n m}^{LG} (x,y,z) & = & C_{n m}^{LG} \(\frac{1}{w}\) e^{-i [k \frac{
r^2}{2 R} +(n+m+1) \psi] - \frac{r^2}{w^2}} \times \nonumber \\
& &  \times e^{-i (n-m) \phi}(-1)^{\mathrm{min}(n,m)} 
\(\frac{r\sqrt{2}}{w}\)^{|n-m|} \times \nonumber \\
& & \ \ \ \times L_{\mathrm{min}(n,m)}^{|n-m|}\(\frac{2 r^2}{w^2}\),
\eea
where $n$, $m$ are arbitrary integers, $R(z)=(z_R^2 + z^2)/z$, 
$\half k w^2(z) = (z_R^2 +z^2)/z_R$,
$\psi (z)=\arctan (z/z_R )$, $z_R$ the Rayleigh range, $H_n (x)$ the
Hermite polynomials and $L_p^l(r)$ the generalized Laguerre polynomials.
The normalization constants are given by $C_{n m}^{HG} = (2/\pi n!
m!)^\half 2^{-\frac{n+m}{2}}$ and $C_{n m}^{LG}= (2/\pi
n! m!)^\half {\mathrm{min}(n,m)}!$. The phase factor $e^{i (n+m+1)
\psi(z)}$ is known
as the Gouy phase, and it is the same only for modes of the same order,
where the order is defined by $N=n+m$. In our implementation we will need
to compensate for this difference in phase.
The LG modes represent states of the photon that have orbital angular
momentum (OAM) $l = |n-m|$~\cite{allen1992a,padgett2002a}. A quantum state 
of the photon can be associated with 
each of these modes~\cite{wunsche2004a}. We will use these states 
in our implementation of the BB84 protocol.

Let
$\{|\psi_i\rangle\}_{i=1}^d$ and $\{|\phi_i\rangle\}_{i=1}^d$ be two
orthonormal bases of a $d$-dimensional Hilbert space. We say that they
are mutually unbiased bases (MUB) if they satisfy
$|\langle \psi_i | \phi_j \rangle |^2 = \frac{1}{d}, \ \ \forall i,j$.
This is the main ingredient of the BB84 protocol. Its main consequence is
that measuring
the state of the system in the ``wrong'' basis gives absolutely no
information about its preparation, since all outcomes are equiprobable.
If we want to implement BB84 with a given physical system as the
information carrier, we need to select at least two bases that are mutually
unbiased. For $d$-dimensional Hilbert spaces, the BB84 protocol can be 
generalized to use more than 2 MUB~\cite{bourennane2001a}, which 
increases the security of the protocol~\cite{bourennane2002a}. 
In $d$ dimensions there 
are at most $(d+1)$ MUB (this bound is known to be tight if $d$ 
is a prime power).

We will now show how to implement the BB84 protocol with photonic
states belonging to a $d$-dimensional Hilbert space.
To simplify the presentation we will neglect for now the effects
of the Gouy phase. We will come back to it after discussing the implementation
of the protocol to show how these phase differences  can be corrected.
First, we 
associate a pure quantum state to each spatial mode of the propagating beam.
The state vector $|m,n\rangle_{HG}$ represents the state of a 
photon propagating
on the spatial mode given by $u_{n m}^{HG}$, while the state 
vector $|m,n\rangle_{LG}$ represents a photon on spatial mode $u_{n m}^{LG}$.
We will consider a $d$-dimensional subspace spanned by the basis ${\cal B}_1 = 
\{ |n\rangle_{HG} \equiv |n,n\rangle_{HG} \}_{n=0}^{d-1}$. 
In the same subspace, we define another  basis ${\cal B}_2 = \{ |\tilde{k}
\rangle_{HG} = \frac{1}{\sqrt{d}} \sum_{n=0}^{d-1} e^{i \frac{2 \pi}{d} k n}
|n\rangle_{HG} \}_{k=0}^{d-1}$. It is easy to
see that these two bases are mutually unbiased.

\begin{figure}[ht]
\centerline{\includegraphics[scale=0.45]{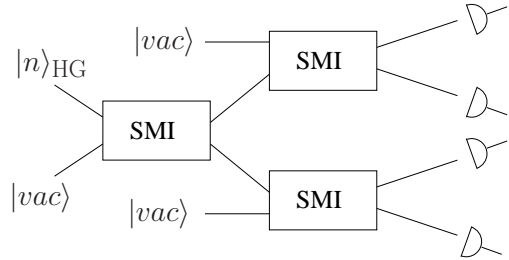}}
\caption{\label{Fig1} Interferometer array used to perform a projective
measurement in the basis ${\cal B}_1$, for $d=4$. Each SMI 
separates photons according to their Hermite-Gauss spatial modes.}
\end{figure}

To implement the BB84 protocol using these two bases we need to 
prescribe how to prepare and how to measure the states in these bases. 
Let us first consider the measurement
problem. For simplicity, let us assume that $d$ is a power of $2$. 
An interferometer that sorts photons according to their HG spatial mode
has been proposed in~\cite{xue2001a}. It consists of cascading Mach-Zender
interferometers, each one containing an appropriate Fractional Fourier
Transform (FRFT)~\cite{lohmann1993a} applied to one of the arms. This 
FRFT is applied by sending the signal through a graded-index (GRIN) rod
of appropriate length, that has a quadratic index profile 
$n(r) = n_0 -n_1 r^2$. This interferometer is called a spatial
modal interleaver (SMI)~\cite{xue2001a}, since the input signal is 
split between the two output ports according to its modal components.
A simple diagram of this cascade of SMI is presented in 
Fig. \ref{Fig1} for the case
$d=4$. 
It is not difficult
to see that the scheme presented in~\cite{xue2001a} can be modified to sort
photons that are prepared in the states of ${\cal B}_1$. Detecting a photon
in one of the arms is equivalent to  a projective
measurement on the basis ${\cal B}_1$.

\begin{figure}[ht]
\centerline{\includegraphics[scale=0.46]{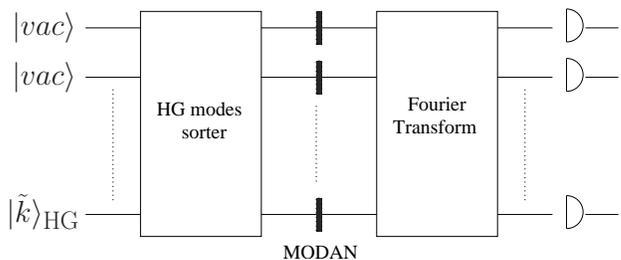}}
\caption{\label{Fig2} Physical setup used to perform a
projective measurement in the basis ${\cal B}_2$. The MODANs 
transform the spatial mode of the photon into a fixed spatial mode
corresponding to the state $|0\rangle_{HG}$.}
\end{figure}

The measurement on the ${\cal B}_2$ basis is a little bit more involved
and requires the use of mode analyzers (MODAN)~\cite{soifer1994a,
duparre1998a} that allow
us to change the spatial mode of a propagating photon. The setup for this
measurement is presented in Fig. \ref{Fig2}. 
The measurement
starts by sending the incoming photon through the same interferometer
used to measure in the ${\cal B}_1$. Let the state of the photon be
$|\tilde{k}\rangle_{HG}$. If $d = 2^s$, the setup has $s$ stages,
with the $j^{th}$ stage consisting of $2^{j-1}$ SMIs. 
The total number of SMIs is then $2^s -1$. Each SMI
requires an extra input mode in the vacuum state, so we should write the
input state as $|\psi\rangle_{input} = |\tilde{k}\rangle_{HG} 
|vac\rangle$, where $|vac\rangle$ represents the vacuum state of
$(2^s -1)$ modes.
After going through the sorter, the state of the system will be 
\be
\label{psip}
|\psi'\rangle = \frac{1}{\sqrt{d}} \sum_{n=0}^{d-1} 
e^{i \frac{2 \pi}{d} k n} |vac\rangle 
\underbrace{|n\rangle_{HG}}_{n^{th} mode} |vac\rangle,
\ee
where the modes on the right-hand side represent the different
output ports of the interferometer. Note that the spatial mode
of the photon is correlated with the output port in which the photon
is present, but we know which port corresponds to each spatial mode.
Now we apply a mode analyzer (MODAN) to each output port such that 
the state of the photon is changed into the state $|0\rangle_{HG}$.
A MODAN that performs this transformation can be implemented
with holographic optical elements~\cite{soifer1994a}. After this step, the
spatial mode of the photon is the same for all the terms in (\ref{psip}),
so to simplify the notation, we will indicate this state by $|1\rangle$,
meaning that one photon is present but its spatial mode
is not relevant. After the MODAN, the state of the system is now
\be
|\psi''\rangle = \frac{1}{\sqrt{d}} \sum_{n=0}^{d-1} 
e^{i \frac{2 \pi}{d} k n} |vac\rangle
\underbrace{|1\rangle}_{n^{th}\, mode} |vac\rangle,
\ee
which represents a superposition, with appropriate phases, of the photon being
in each one of the output ports of the HG mode sorter.  
Now we can simply
apply a Fourier transform to these $d$ modes,
which can be accomplished by using linear optical elements such
as mirrors, beam splitters and phase shifters~\cite{reck1994a}. 
After this step, the state of the system is given by
\be
|\psi\rangle_{output} = |vac\rangle 
\underbrace{|1\rangle}_{k^{th}\, mode} |vac\rangle.
\ee
By placing photo detectors in the output modes of the
Fourier transform (FT) device we can measure the value of $k$, which
accomplishes a projective measurement in the basis ${\cal B}_2$.
The step that employs the MODAN is an example of the quantum erasure
effect: by ``erasing'' the spatial mode information from the photon
state, 
we erase the ``which path'' information, which allows us to extract the
value of $k$ from the relative phases by using a Fourier transform.

The preparation of  states in the two bases can also
be solved with the help of MODANs. To prepare a state from
${\cal B}_1$ we will assume that we have a single photon gun the
produces a photon in the state $|0\rangle_{HG}$ which is just the
usual lowest order Gaussian mode. By sending the photon through
an appropriate MODAN, we can in principle transform the spatial
mode, thus preparing any other state in ${\cal B}_1$. To prepare 
a state in ${\cal B}_2$, we can just run the device we use to 
perform the projective measurement in the basis ${\cal B}_2$ backwards. 
By appropriately 
selecting one of the ``output'' ports of the FT device 
to send a photon through, we can choose the value of $k$ for the
state that will be output at the other end.

The implementation we have presented so far still requires the sender and
the receiver to align their reference systems, because the HG spatial
modes given by (\ref{modes}) clearly single out a direction on the plane
perpendicular to the direction of propagation. However, we can get rid
of this requirement by using the LG modes for the transmission portion of the
protocol. This can be accomplished by using a modal 
converter~\cite{beijersbergen1993a,padgett2002a} consisting 
of two cylindrical lenses, that 
transform the spatial mode $u_{n m}^{HG} (x,y,z)$ into 
$u_{n m}^{LG} (x,y,z)$ and vice versa. 

Our implementation of a $d$-dimensional BB84 protocol with photonic
spatial modes will be the following. First, Alice randomly chooses which one
of the bases ${\cal B}_1$ and ${\cal B}_2$ she will use
to send the information. Then she randomly chooses a number $k$
between $0$ and $d-1$. If she chose the basis ${\cal B}_1$,
she prepares the state $|k\rangle_{HG}$ which is associated
with the Hermite-Gauss spatial mode $u_{k k}^{HG} (x,y,z)$. Then 
she runs it through a modal converter, that transforms
into the state $|k\rangle_{LG}$, associated with the Laguerre-Gauss 
spatial mode $u_{k k}^{LG} (x,y,z)$, and then she sends the state to Bob.
From (\ref{modesLG}) we can see that the state $|k\rangle_{LG}$ has
no dependence on the azimuthal angle $\phi$ (i.e., it has zero orbital angular
momentum), and hence \emph{it is invariant under rotations
about the propagation direction}. On the other hand, if Alice
choses the basis ${\cal B}_2$, she prepares the state 
$|\tilde{k}\rangle_{HG} = \frac{1}{\sqrt{d}} \sum_{n=0}^{d-1} 
e^{i \frac{2 \pi}{d} k n} |n\rangle_{HG}$. She runs it through the modal
converter, thus obtaining the state $|\tilde{k}\rangle_{LG} = 
\frac{1}{\sqrt{d}} \sum_{n=0}^{d-1} e^{i \frac{2 \pi}{d} k n} |n\rangle_{LG}$.
As before, we can see each state in this superposition is 
invariant under rotations about the propagation direction. Then, 
\emph{the state $|\tilde{k}\rangle_{LG}$ itself is invariant under
these rotations}. 

On the other end, Bob sends the photon he receives through a modal
converter, that transforms the LG states back into HG states.
He then randomly chooses in which basis he wants to measure the photon,
and performs the corresponding measurement. The rest
of the protocol is the usual BB84. Note that once the photon is sent
through the modal converter, a direction in the plane perpendicular to
the propagation direction is singled out, since the cylindrical lenses
have a preferred direction in that plane. The interferometer employed
to do the measurement also has a preferred direction, and this
direction must be consistent with the one chosen by the 
modal converter. However, since the incoming state of the
photon is rotationally invariant on that plane, the direction
singled out by the cylindrical lenses is \emph{completely arbitrary}.
By encoding the information in rotationally invariant states
we have effectively decoupled the alignment between the preparation
and measuring devices. 

We come back now to the problem of  the phase differences between
propagating modes of different orders introduced by the Gouy phase.
This problem does not affect the preparation and measurement of states
in the ${\cal B}_1$ basis, since it only adds an overall phase.
For states in the ${\cal B}_2$ basis however, the relative phases of
different spatial modes are crucial. If we consider the devices that 
prepare and measure states in the ${\cal B}_2$ basis, since they 
separate the spatial modes by sending them through different arms of
an interferometric device, it is clear that we can compensate any
extra phase factors by applying appropriate phase shifts. The problem
of dephasing during transmission, where the state is encoded as a 
superposition of LG modes, can be also fixed by appropriate phase
shifters applied within the measuring device. To see this, note that 
that the $z$ dependence of the Gouy phase comes from the function 
$\psi(z) = \arctan (z/z_R)$, where $z_R$ is the Rayleigh range. 
Thus, this mode-dependent phase shift can be compensated if we 
know the distance between the sender and the receiver. Also, this
problem simplifies in the limit
in which this distance is much larger than the Rayleigh range. Since the
modes we use have order $2n$, the
phase shift in this case takes only the values 
$\frac{\pi}{2}$  and $\frac{3\pi}{2}$,
depending on whether $n$ is even or odd respectively, 
which can be compensated by phase shifters inserted in the measuring device.

In this implementation, the transmitted state of the photon was
encoded in a subspace of spatial modes that have zero angular
momentum. We can also implement this protocol by encoding in
a subspace with some definite value $l$ of orbital angular momentum.
These states are associated with the LG spatial modes 
$u_{(n+l) n}^{LG} (x,y,z)$. Even though these modes are not invariant
under rotations about the propagation axis, they only pick 
up a phase factor $e^{i l \phi_0}$, where $\phi_0$ 
is the angle of rotation. An overall phase factor is irrelevant for a
pure state. Moreover, any linear superposition of these states
will also pick up the same global phase factor. We can use
this subspace of states with $l$ orbital angular momentum to implement
our protocol, and this implementation still does not require
alignment between Alice and Bob. Since there are simple
ways of sorting states with different orbital angular 
momentum~\cite{wei2003a,leach2002a}, we could simultaneously 
implement QKD on the same channel 
using different values of OAM (orbital angular momentum multiplexing).

The alignment independence of our implementation can be taken 
a step further by noting that everything we discussed holds true \emph{even 
when the angle between Alice and Bob's reference frames is a function
of time. The protocol still allows for QKD when the sender and
receiver are rotating with respect to each other about the axis of 
propagation of the signal}. However, we should note that this property
holds only when we use states with zero OAM for
transmission. States with nonzero OAM undergo a frequency
shift due to the rotation~\cite{nienhuis1996a,courtial1998a}, which
affects the performance of the interferometers used in their
measurement. However, if the interferometers are calibrated to
compensate for this shift (which in practice implies
knowing the relative angular velocity between Alice and Bob),
we can still use higher OAM states for QKD.

It is also worth noting that we could implement the BB84 protocol
by encoding the information in photon OAM states and their superpositions,
that can be analyzed using the techniques described in~\cite{wei2003a}.
In this case however, the protocol requires alignment between Alice
and Bob. But we can take advantage of this alignment
to double the size of the Hilbert space in which the information 
is encoded, by combining both OAM and polarization degrees of freedom.
A detailed description of this implementation will be presented
elsewhere~\cite{huver2004a}.

In this letter, we introduced a novel implementation of the BB84
quantum key distribution protocol, that encodes information in
the spatial modes of propagating photons. This has two main
advantages. First, by employing a $d$-dimensional Hilbert
space we can increase the key generation rate by increasing
the bits per photon that can be sent. This gain is only logarithmic
in $d$, but doubling or tripling the key generation rate may be feasible
with current technology. On the other hand, 
this implementation does not require
reference frame alignment between Alice and Bob, making it
particularly appealing for key distribution between moving
stations, such as satellites in space. 

I would like to thank Deborah Jackson for suggesting this problem
to me, and Jonathan Dowling for his encouragement
and useful suggestions to improve this manuscript. This work was
carried out at the Jet Propulsion Laboratory, California Institute 
of Technology, under a contract
with the National Aeronautics and Space Administration (NASA).
This work was supported by the National Research Council and
NASA, Code Y.

\bibliographystyle{prsty}
\bibliography{Cav}

\begin{thebibliography}{10}

\bibitem{nielsen2000}
M.~N. Nielsen and I.~L. Chuang, {\em Quantum computation and quantum
  information} (Cambridge University Press, Cambridge, 2000).

\bibitem{molina2002a}
G. Molina-{T}erriza, J.~P. Torres, and L. Torner, Phys. Rev. Lett. {\bf 88},
  013601  (2002).

\bibitem{mair2001a}
A. Mair, A. Vaziri, G. Weihs, and A. Zeilinger, Nature {\bf 412},  313  (2001).

\bibitem{langford2004a}
N.~K. Langford {\it et~al.}, Phys. Rev. Lett. {\bf 93},  053601  (2004).

\bibitem{bartlett2003a}
S.~D. Bartlett, T. Rudolph, and R.~W. Spekkens, Phys. Rev. Lett. {\bf 91},
  027901  (2003).

\bibitem{bennett1984a}
C.~H. Bennett and G. Brassard, {\em Proceedingd of {IEEE} {I}nternational
  {C}onference on {C}omputers, {S}ystems and {S}ignal {P}rocessing} (IEEE, New
  York, 1984), pp.\ 175--179.

\bibitem{beijersbergen1993a}
M.~W. Beijersbergen, L. Allen, H.~E. L.~O. van~der Veen, and J.~P. Woerdman,
  Optics Communications {\bf 96},  123  (1993).

\bibitem{wunsche2004a}
A. W\"{u}nsche, J. Opt. B: Quantum Semiclass. Opt. {\bf 6},  S47  (2004).

\bibitem{abramochkin2004a}
E.~G. Abramochkin and V.~G. Volostnikov, J. Opt. B: Quantum Semiclass. Opt.
  {\bf 6},  S157  (2004).

\bibitem{allen1992a}
L. Allen, M.~W. Beijersbergen, R.~J.~C. Spreeuw, and J.~P. Woerdman, Phys. Rev.
  A {\bf 45},  8185  (1992).

\bibitem{padgett2002a}
M.~J. Padgett and L. Allen, J. Opt. B: Quantum Semiclass. Opt. {\bf 4},  S17
  (2002).

\bibitem{bourennane2001a}
M. Bourennane, A. Karlsson, and G. Bj\"{o}rk, Phys. Rev. A {\bf 64},  012306
  (2001).

\bibitem{bourennane2002a}
M. Bourennane {\it et~al.}, J. Phys. A {\bf 35},  10065  (2002).

\bibitem{xue2001a}
X. Xue, H. Wei, and A.~G. Kirk, Optics Letters {\bf 26},  1746  (2001).

\bibitem{lohmann1993a}
A.~W. Lohmann, J. Opt. Soc. Am. A {\bf 10},  2181  (1993).

\bibitem{soifer1994a}
V. Soifer and M. Golub, {\em Laser {B}eam {M}ode {S}election by {C}omputer
  {G}enerated {H}olograms} (CRC Press, Boca Raton, 1994).

\bibitem{duparre1998a}
M. Duparre {\it et~al.}, Proc. SPIE {\bf 3291},  104  (1998).

\bibitem{reck1994a}
M. Reck, A. Zeilinger, H.~J. Bernstein, and P. Bertani, Phys. Rev. Lett. {\bf
  73},  58  (1994).

\bibitem{wei2003a}
H. Wei {\it et~al.}, Optics Communications {\bf 223},  117  (2003).

\bibitem{leach2002a}
J. Leach {\it et~al.}, Phys. Rev. Lett. {\bf 88},  257901  (2002).

\bibitem{nienhuis1996a}
G. Nienhuis, Optics Communications {\bf 132},  8  (1996).

\bibitem{courtial1998a}
J. Courtial {\it et~al.}, Phys. Rev. Lett. {\bf 80},  3217  (1998).

\bibitem{huver2004a}
S. Huver and F. M. Spedalieri, in preparation.

\end{thebibliography}

\end{document}